# Robust Control of Nanoscale Drug Delivery System in Atherosclerosis: A Mathematical Approach

**Alireza Rowhanimanesh[1,2] and Mohammad-R. Akbarzadeh-T.[2] ***

[1] Department of Electrical Engineering, University of Neyshabur, Neyshabur, Iran.
[2] Department of Electrical Engineering, Center of Excellence on Soft Computing and Intelligent Information Processing, Ferdowsi University of Mashhad, Mashhad, Iran. * e-mail: akbazar@um.ac.ir

**Abstract.** This paper proposes a mathematical approach for robust control of a nanoscale drug delivery system in treatment of atherosclerosis. First, a new nonlinear lumped model is introduced for mass transport in the arterial wall, and its accuracy is evaluated in comparison with the original distributed-parameter model. Then, based on the notion of sliding-mode control, an abstract model is designed for a smart drug delivery nanoparticle. In contrast to the competing strategies on nanorobotics, the proposed nanoparticles carry simpler hardware to penetrate the interior arterial wall and become more technologically feasible. Finally, from this lumped model and the nonlinear control theory, the overall system's stability is mathematically proven in the presence of uncertainty. Simulation results on a well-known model, and comparisons with earlier benchmark approaches, reveals that even when the LDL concentration in the lumen is high, the proposed nanoscale drug delivery system successfully reduces the drug consumption levels by as much as 16% and the LDL level in the Endothelium, Intima, Internal Elastic Layer (IEL) and Media layers of an unhealthy arterial wall by as much as 14.6%, 50.5%, 51.8%, and 64.4%, respectively.

## 1. Introduction

Atherosclerosis, or hardening of the arteries, is a leading cause of death in humans. Medical research has shown that an abnormally high accumulation of LDL (low-density lipoprotein) macromolecules in the arterial wall plays a critical role in the initiation and development of atherosclerotic plaques [1-3]. In contrast to the existing methods such as global drug delivery, angioplasty, and open surgery, targeted drug delivery by Drug-Encapsulated Nanoparticles (DENPs) has recently received significant attention as a promising non-invasive approach to prevent and treat atherosclerosis. DENPs' selective drug delivery reduces the unwanted side effects of the treatment while also increasing its effectiveness. This rising promise has also come with considerable advances in nanomedicine and the manufacturing technology of DENPs.

Nevertheless, there remain only a few, if any, theoretical developments in this realm. Besides providing rigorous mathematical proof of performance, such as the stability of such systems, theoretical developments could pave the way for realizing less complicated and more effective nanoparticle designs.

Several recent works address nanomedicine's application to the prevention, diagnosis, and treatment of atherosclerosis [4-6]. Advanced nanostructures are widely employed for therapeutic tasks in atherosclerosis, such as multifunctional nanoparticles with combined diagnostic and therapeutic capacities as theranostic nanosystems [7], engineered nanoparticles [8], carbon nanotubes [9], targeted nanocarriers [10], mechanoresponsive nanoplatforms [11], and stimuli-responsive nanoagents [12]. Nanoagents are also used for diagnostic goals, particularly for atherosclerosis imaging [13]. In addition to experimental and clinical studies, many investigations are performed on mathematical modeling and computer numerical simulation of atherosclerosis treatment by nanoagents [14-15].

The basic notion of Autonomous DENPs (ADENPs) was first introduced by authors in 2013 [16-17]. In contrast to theoretical nanorobots that are not experimentally feasible by today's technology, ADENPs aim for a technologically simpler and yet autonomous solution to drug delivery by emphasizing a massive number of the nanoparticles, in which the intelligence rises from the synergism of their swarm architecture. With this general concept in mind, the authors proposed Proportional DENPs (PDENPs) [17] for the prevention of atherosclerosis by reducing the LDL concentration in the arterial wall's interior. PDENPs are drug encapsulated nanoparticles that are equipped with two sensing devices for LDL and drug level concentrations. In contrast to the existing targeted DENPs that usually target the surface proteins of the plaque, PDENPs directly sense the LDL level in the arterial wall. Hence, they can diagnose abnormal LDL accumulation before plaque formation and prevent the plaque's critical growth. More recently, authors [19] proposed a stigmergic approach to the cooperation of two different ADENP constructs for swarm control of LDL concentration in the arterial wall. Through this simple proportional feedback and stigmergic approach, authors showed that there can be viable alternatives to the prevention of atherosclerosis within the realm of ADENPs.

The point of departure in the present study is introducing the principles of systems theory and mathematical analysis in the design of nanoparticles. Besides providing rigorous mathematical results, such analytical approach may help determine where the design of nanoparticles can be simplified. Hence, it can also lead to simpler and more realizable nanoparticles. Here, we propose a sliding-mode approach to proportional drug-encapsulated nanoparticles (SPDENPs) based on the mathematical principles of sliding mode control theory. We also propose a new nonlinear lumped mathematical model for mass transport in the arterial wall and use it for the analytical design of SPDENPs. Similar to the earlier works of the authors, the goal of SPDENPs is to quickly reduce LDL level in the interior of the arterial wall, not in the blood (lumen), to prevent critical growth of atherosclerotic plaques. Thus, SPDENPs are mainly designed for patients with very high LDL levels who correspond to the highest cardiovascular disease rates.

There are several noteworthy features of the proposed SPDENPs. Firstly, using the analytical approach to design and the principles of robust control theory, the stability of SPDENPs is mathematically proven in the presence of uncertainty. In contrast, our earlier methods [18-19] were designed through ad hoc procedures using numerical simulations without any mathematical proof. Secondly, SPDENPs have a simpler structure. PDENPs sense both LDL and drug concentrations and use piece-wise linear controllers to reduce LDL level in the arterial wall. Whereas SPDENPs sense only LDL concentration and its rate of change, and benefit from a nonlinear-proportional control structure. Since PDENPs take feedback from the drug, they can reduce unwanted drug side effects. In contrast, SPDENPs do not require drug concentration sensors, leading to a simpler design. The same can be said about the added complexity of the approach in [19] that is based on pheromone sensing and stigmergic cooperation between two types of ADENPs. Thirdly, simulation results show that lower drug consumption and improved LDL reduction is obtained by SPDENPs. These simulations also illustrate that SPDENPs are generally more robust than PDENPs, i.e., they maintain better performance in the presence of uncertainty.

It should be mentioned that similar to our earlier works, we consider each nanoparticle as a system, and our insight to SPDENP is abstract and mathematical. Inspired from systems biology, the whole is greater than the sum of its parts, and this computational modeling approach helps us to holistically analyze complex interactions and emerging behaviors of biological systems using in-silico study. We hope that the present paper could draw an outline of a new practical approach to targeted drug delivery that is rooted in a mathematical foundation. This could open new possibilities for introducing mathematical rigor in this highly applied field.

The rest of the paper is organized as follows. In Section 2, after a brief review of the mathematical modeling of mass transport in the arterial wall, a new nonlinear lumped model is introduced, and its accuracy is demonstrated through simulations. In Section 3, the proposed control approach is designed based on mathematical proof, and the structure of the corresponding SPDENPs is explained. Simulation results on the original model and comparisons with PDENPs are then provided in Section 4. Finally, conclusions are offered in Section 5.

## 2. Mathematical Modeling

### *2.1. Mass Transport in the Arterial Wall*

The present study utilizes the four-layer (i.e. Endothelium, Intima, IEL, and Media) model of the arterial wall as defined in [1]. Fig.1-a shows the thickness of each wall layer and boundary conditions. The details of the model and the values of physiological parameters are presented in [18]. For completeness, the governing equations in each wall layer are as follows:

$$\frac{\partial C_{LDL}}{\partial t} = -(1-\sigma^{l}_{fLDL}).V_{filt}.\nabla C_{LDL} + D^{l}_{LDL}.\nabla^2 C_{LDL} - k^{l}_{rLDL}.C_{LDL} - R_y(C_{LDL}, C_{drug}) \tag{1}$$

$$\frac{\partial C_{drug}}{\partial t} = -(1-\sigma^{l}_{fdrug}).V_{filt}.\nabla C_{drug} + D^{l}_{drug}.\nabla^2 C_{drug} - k_{rdrug}.C_{drug} - R_z(C_{LDL}, C_{drug}) + m.C_{DENP}.u_{drug} \tag{2}$$

$$\frac{\partial C_{DENP}}{\partial t} = -(1-\sigma^{l}_{fDENP}).V_{filt}.\nabla C_{DENP} + D^{l}_{DENP}.\nabla^2 C_{DENP} - k_{rDENP}.C_{DENP} \tag{3}$$

where $l$ is the layer index, $V_{filt}$ is the filtration velocity, $m$ is the mass of drug molecules, $k_{rdrug}$ and $k_{rDENP}$ are the reaction coefficients of drug and SPDENP with the environment, respectively. The reaction between LDL and drug are modeled by $R_y$ and $R_z$, respectively. Also, $C_{LDL}$, $\sigma^{l}_{fLDL}$, $D^{l}_{LDL}$ and $k^{l}_{rLDL}$ are the concentration, filtration reflection, effective diffusivity, and reaction coefficients, respectively, for LDL in the $l^{th}$ layer. Without a loss of generality, a one-dimensional model of equations (1)-(3) is used here, in which all signals are just spatially dependent on *r* axis [18]. Similar notations are used for the drug and SPDENPs. $u_{drug}$ is the drug release rate (molecules per second) by each SPDENP. SPDENPs and drug molecules are assumed to be spherical particles of 100 nm and 4 nm in diameter, respectively. Previous experimental studies in the literature have demonstrated that this assumed size for SPDENPs is reasonable [18].

### *2.2. The Proposed Lumped State Space Model*

A lumped approximation of (1)-(3) is obtained by spatial discretization of the model of Fig.1-a over the radial axis ($r$) with a mesh size of $h_r$ (100 nm in this study) as displayed in Fig.1-b. The equations of the lumped model are derived as follows:

$$\frac{dy_i}{dt} = c_{1i}^y . y_{i-1} + c_{2i}^y . y_i + c_{3i}^y . y_{i+1} - R_y(y_i, z_i) \tag{4}$$

$$\frac{dz_i}{dt} = c_{1i}^z . z_{i-1} + c_{2i}^z . z_i + c_{3i}^z . z_{i+1} - R_z(y_i, z_i) + m.n_i.u_i \tag{5}$$

$$\frac{dn_i}{dt} = c_{1i}^n . n_{i-1} + c_{2i}^n . n_i + c_{3i}^n . n_{i+1} \tag{6}$$

where $i = 1,...,q$ is the index of the finite element node in Fig.1-b, where $q = \frac{1}{h_r}\sum_{k=1}^{4} L_k$ and $L_k$ is the length of each layer. For notational simplicity, $y_i$, $z_i$ and $n_i$ are $C_{LDL}$, $C_{drug}$ and $C_{DENP}$ at the $i^{th}$ finite element node, respectively. Also $y_0 = \hat{C}_{LDL}$, $z_0 = \hat{C}_{drug}$, and $n_0 = \hat{C}_{DENP}$ are the concentrations of LDL, drug, and DENP in the lumen, respectively.

At each finite element node $i$, we then have:

$$c_{1i}^y = \frac{(1-\sigma_{fLDL}^l)V_{filt}}{h_r} + \frac{D_{LDL}^l}{h_r^2}, c_{2i}^y = -\frac{(1-\sigma_{fLDL}^l)V_{filt}}{h_r} - 2\frac{D_{LDL}^l}{h_r^2} - k_{rLDL}^l\ c_{3i}^y = \frac{D_{LDL}^l}{h_r^2}, c_{1i}^z = -\frac{(1-\sigma_{fdrug}^l)V_{filt}}{h_r} + \frac{D_{drug}^l}{h_r^2}, c_{3i}^z = \frac{D_{drug}^l}{h_r^2}$$

$$c_{2i}^z = -\frac{(1-\sigma_{fdrug}^l)V_{filt}}{h_r} - 2\frac{D_{drug}^l}{h_r^2} - k_{rdrug}^l, c_{1i}^n = \frac{(1-\sigma_{fDENP}^l)V_{filt}}{h_r} + \frac{D_{DENP}^l}{h_r^2}\ c_{2i}^n = -\frac{(1-\sigma_{fDENP}^l)V_{filt}}{h_r} - 2\frac{D_{DENP}^l}{h_r^2} - k_{rDENP}^l, c_{3i}^n = \frac{D_{DENP}^l}{h_r^2}$$

where $l$ is the layer index. At the boundary nodes, the above equations are similarly available, except for the following nodes:

$$\to i = \frac{L_1}{h_r} + 1: \quad c_{1i}^y = \frac{(1-\sigma_{fLDL}^1)V_{filt}}{h_r} + \frac{D_{LDL}^1}{h_r^2}, \quad c_{2i}^y = -\frac{(1-\sigma_{fLDL}^2)V_{filt}}{h_r} - \frac{(D_{LDL}^1 + D_{LDL}^2)}{h_r^2} - k_{rLDL}^2, \quad c_{3i}^y = \frac{D_{LDL}^2}{h_r^2}$$

$$\to i = \frac{L_1 + L_2}{h_r}: \quad c_{1i}^y = \frac{(1-\sigma_{fLDL}^2)V_{filt}}{h_r} + \frac{D_{LDL}^2}{h_r^2}, \quad c_{2i}^y = -\frac{(1-\sigma_{fLDL}^2)V_{filt}}{h_r} - \frac{(D_{LDL}^2 + D_{LDL}^3)}{h_r^2} - k_{rLDL}^2, \quad c_{3i}^y = \frac{D_{LDL}^3}{h_r^2}$$

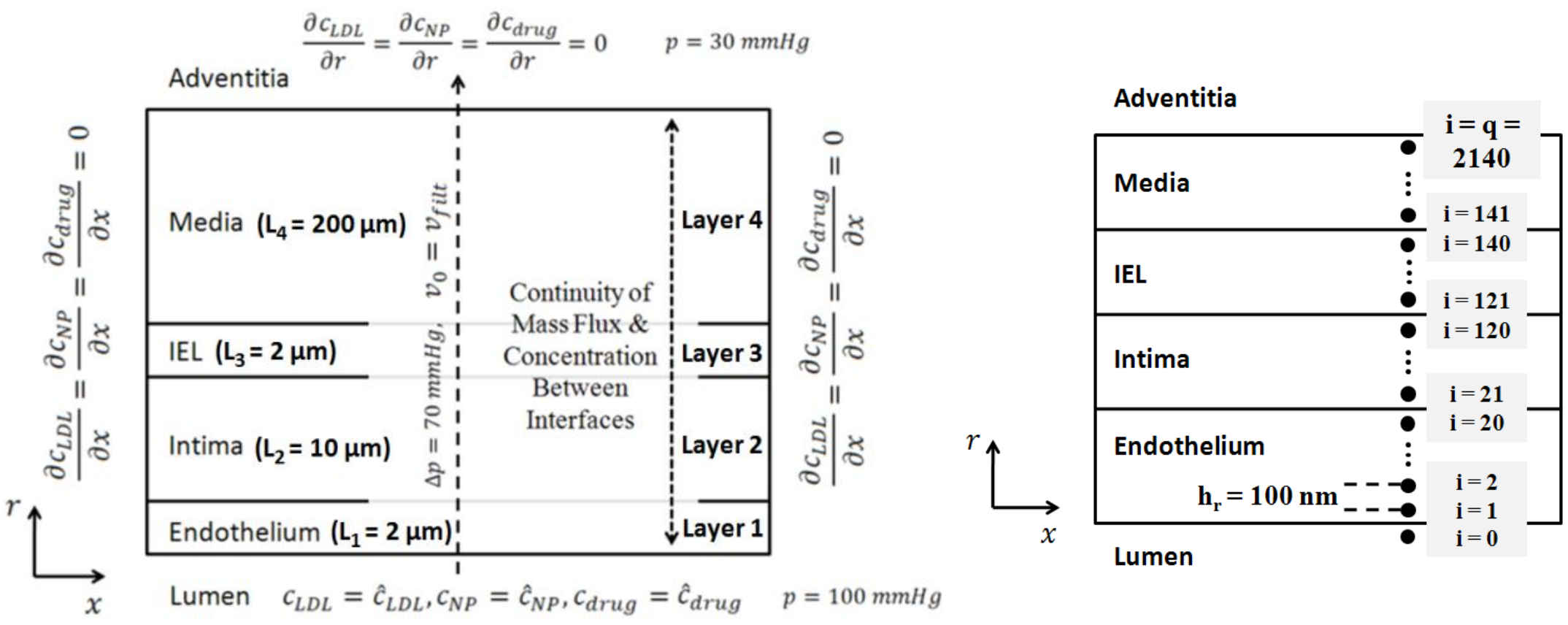

(a) (b)

**Figure 1 | a,** The boundary conditions used in the numerical simulation. **b,** Discretization of the model of the arterial wall over the radial axis.

$$\rightarrow i = \frac{L_1 + L_2}{h_r} + 1: \quad c_{1i}^y = \frac{(1-\sigma_{fLDL}^2)V_{filt}}{h_r} + \frac{D_{LDL}^3}{h_r^{\ 2}}, \quad c_{2i}^y = -\frac{(1-\sigma_{fLDL}^3)V_{filt}}{h_r} - 2\frac{D_{LDL}^3}{h_r^{\ 2}} - k_{rLDL}^3, \quad c_{3i}^y = \frac{D_{LDL}^3}{h_r^{\ 2}}$$

$$\rightarrow i = \frac{L_1 + L_2 + L_3}{h_r} + 1: \quad c_{1i}^y = \frac{(1-\sigma_{fLDL}^3)V_{filt}}{h_r} + \frac{D_{LDL}^3}{h_r^{\ 2}}, \quad c_{2i}^y = -\frac{(1-\sigma_{fLDL}^4)V_{filt}}{h_r} - \frac{(D_{LDL}^3 + D_{LDL}^4)}{h_r^{\ 2}} - k_{rLDL}^4, \quad c_{3i}^y = \frac{D_{LDL}^4}{h_r^{\ 2}}$$

$$\rightarrow i = q: \quad c_{1i}^y = \frac{(1-\sigma_{fLDL}^4)V_{filt}}{h_r} + \frac{D_{LDL}^4}{h_r^{\ 2}}, \quad c_{2i}^y = -\frac{D_{LDL}^4}{h_r^{\ 2}} - k_{rLDL}^4$$

The parameters $c_{1i}^z, c_{2i}^z, c_{3i}^z, c_{1i}^n, c_{2i}^n$ and $c_{3i}^n$ are defined similarly. According to the physiological and chemical characteristics, at all finite element nodes, the following inequalities hold:

$$c_{1i}^y, c_{3i}^y, c_{1i}^z, c_{3i}^z, c_{1i}^n, c_{3i}^n \geq 0, c_{2i}^y, c_{2i}^z, c_{2i}^n \leq 0, y_i, z_i, n_i, y_0, z_0, n_0 \geq 0\,. \tag{7}$$

Also, $R_y(y_i, z_i)$ and $R_z(y_i, z_i)$ are non-negative, increasing, and globally Lipschitz continuous over $R^2 - \{\infty\}$, and $R_y(0, z_i) = R_y(y_i, 0) = R_y(0,0) = R_z(0, z_i) = R_z(y_i, 0) = R_z(0,0) = 0$. These functions are usually modeled as $\alpha y_i^{\beta_y} z_i^{\beta_z}$, with $\alpha > 0$, and $\beta y$ and $\beta z$ being positive constant integers. Equations (4) and (5) can be rewritten in the state space form as follows:

$$\dot{X}(t) = AX(t) - R(X(t)) + L(t) + B_u(t)U(t) \tag{8}$$

where $X(t) = [y_1(t), ..., y_q(t), z_1(t), ..., z_q(t)]^T$ is the state vector,

$$
A=\begin{bmatrix}
c_{21}^{y} & c_{31}^{y} & 0 & \dots & 0 & & & & & \\
c_{12}^{y} & \ddots & \ddots & \ddots & \vdots & & & & & \\
0 & \ddots & \ddots & \ddots & 0 & & & 0 & & \\
\vdots & \ddots & \ddots & \ddots & c_{3(q-1)}^{y} & & & & & \\
0 & \dots & 0 & c_{1q}^{y} & c_{2q}^{y} & & & & & \\
 & & & & & c_{21}^{z} & c_{31}^{z} & 0 & \dots & 0 \\
 & & & & & c_{12}^{z} & \ddots & \ddots & \ddots & \vdots \\
 & & 0 & & & 0 & \ddots & \ddots & \ddots & 0 \\
 & & & & & \vdots & \ddots & \ddots & \ddots & c_{3(q-1)}^{z} \\
 & & & & & 0 & \dots & 0 & c_{1q}^{z} & c_{2q}^{z}
\end{bmatrix},
$$

$U(t)=[u_1(t),...,u_q(t)]^T$ is the control input,

$B_u(t)=[0_{q\times q} \quad m.diag([n_1,...,n_q])]^T$,

$L(t)=[c_{11}^{y}\hat{C}_{LDL}(t) \quad 0_{1\times(q-1)} \quad c_{11}^{z}\hat{C}_{drug}(t) \quad 0_{1\times(q-1)}]^T$,

$R(X(t))=[R_y(y_1,z_1),...,R_y(y_q,z_q),R_z(y_1,z_1),...,R_z(y_q,z_q)]^T$. According to (6), $N(t)=[n_1(t),...,n_q(t)]^T$ is independent from $X(t)$, and thus it can be described by a separate LTI (linear time invariant) state equation as follows:

$$\dot{N}(t)=A_nN(t)+B_n\hat{C}_{DENP}(t) \tag{9}$$

where $B_n=[c_{11}^{n},0_{1\times(q-1)}]^T$, and

$$
A_n=\begin{bmatrix}
c_{21}^{n} & c_{31}^{n} & 0 & \dots & 0 \\
c_{12}^{n} & \ddots & \ddots & \ddots & \vdots \\
0 & \ddots & \ddots & \ddots & 0 \\
\vdots & \ddots & \ddots & \ddots & c_{3(q-1)}^{n} \\
0 & \dots & 0 & c_{1q}^{n} & c_{2q}^{n}
\end{bmatrix}.
$$

### *2.3. Model Validation*

In this section, we compare the proposed lumped model in equations (8)-(9) with the original distributed model in equations (1)-(3) through numerical simulation in MATLAB [20]. The number of finite element nodes *q* is 2140 due to the mesh size of 100 nm for wall thickness of 214 μm (the four layers of Fig.1-a). Also, a time step of 10 seconds is considered for both models. More details about the values of other parameters are provided in Section 4. It is assumed here that the proposed SPDENPs are administrated to a patient who has a very high LDL level. Thus, the LDL concentration in the lumen, $\hat{C}_{LDL}$, is set to be 200 *mg/dL* [21]. We further presume that the concentration of SPDENP and drug are $\hat{C}_{SPDENP}=200\ \mu g/mL$ and 0 in the lumen, respectively.

Both models are initialized as follows: the LDL concentration (state) is set to the uncontrolled (without drug) LDL profile (the equilibrium point of (1) when $C_{drug}$ is zero), and SPDENP and drug concentrations are set to zero. Inspired by the systems identification theory, we have used a uniformly distributed independent random process (in the range of 0 and 200 molecules per second) for each $u_i$ ($i=1:q$) in (2) and (8) as the test input vector. For illustration, $u_1$ is displayed in Fig.2-d.

We consider the effect of SPDENPs on the LDL level of the interior of the arterial wall over six hours for both the proposed and original models. Fig.2 depicts the final profiles of LDL, drug and SPNENP concentrations after six hours. Parts a and c show that the proposed lumped model can adequately approximate the original distributed model in LDL and SPDENP concentrations. The approximation error of drug concentration is comparatively larger, especially in the Media layer. This may be because of the simultaneous effects (and approximation errors) of both LDL and SPDENP concentrations in the drug equations (2) and (8). Also, due to the direction of fluid velocity ($V_{filt}$), from lumen towards Media (i.e. only the radial component of velocity is considered to be nonzero [14]), and the boundary condition of drug concentration in lumen (that is zero against the boundary condition of SPDENP and LDL), the concentration of drug and consequently the approximation error is large in Media.

In this simulation, the mean absolute approximation error of LDL, drug, and SPDENP concentrations in all of the arterial wall's finite element nodes and over six hours are 0.82349 mg/dL, 3.084 μg/mL, and 0.38684 μg/mL, respectively. These approximation errors are equal to 0.42%, 12.83%, and 0.22% of the maximum value of these signals over six hours. The results demonstrate that the proposed lumped model can be used as an adequate approximation of the original distributed model of mass transport in the arterial wall.

In the next section, this lumped model (equations (8) and (9)) is then employed for controller design. It should be noted that the above treatment of mass transport of the arterial wall aims to provide a new venue for an abstract and mathematical approach to the treatment of atherosclerosis. In contrast to the current state-of-the-art that is purely empirical, such an approach could potentially reduce the costly and time-consuming laboratory process while also providing mathematically verifiable solutions. There are a considerable number of other interesting aspects to arterial failure that can benefit from this scheme, such as fluid pressure [22], fluid-structure interaction [23], and the effect of gender-related geometrical characteristics of the Aorta-Iliac Bifurcation [24]. We believe this study may also serve as a mathematical basis for these and other similar effects.

## 3. Controller Design

### *3.1. Systems Equations*

The task of the SPDENP swarm here is reducing the LDL level $Y(t)=[y_1(t),...,y_q(t)]^T$ by manipulating the drug release rate $U(t)=[u_1(t),...,u_q(t)]^T$. From the viewpoint of a SPDENP at the $i^{th}$ finite element node, the control system's input and output are $u_i(t)$ and $y_i(t)$, respectively. We derive the governing equations between $y_i(t)$ and $u_i(t)$ by differentiating (4) with respect to $t$ as follows:

$$\frac{d^2y_i}{dt^2}=c_{1i}^y\frac{dy_{i-1}}{dt}+c_{2i}^y\frac{dy_i}{dt}+c_{3i}^y\frac{dy_{i+1}}{dt}-\frac{dy_i}{dt}\frac{\partial R_y(y_i,z_i)}{\partial y_i}-\frac{dz_i}{dt}\frac{\partial R_y(y_i,z_i)}{\partial z_i} \tag{10}$$

For simplicity, $g_i$ is defined, and (4) is rewritten as follows:

$$\frac{dy_i}{dt}=g_i(y_{i-1},y_i,y_{i+1},z_i)=c_{1i}^y y_{i-1}+c_{2i}^y y_i+c_{3i}^y y_{i+1}-R_y(y_i,z_i) \tag{11}$$

Using (11), $dy_{i-1}/dt$ and $dy_{i+1}/dt$ can be presented as:

$$\frac{dy_{i-1}}{dt}=g_{i-1}(y_{i-2},y_{i-1},y_i,z_{i-1}) \tag{12}$$

$$\frac{dy_{i+1}}{dt}=g_{i+1}(y_i,y_{i+1},y_{i+2},z_{i+1}) \tag{13}$$

Using Equations (5) and (11)-(13) in (9) leads to the below governing relation between $y_i$ and $u_i$:

$$\frac{d^2y_i}{dt^2}=f_i(y_{i-2},y_{i-1},y_i,y_{i+1},y_{i+2},z_{i-1},z_i,z_{i+1})-b(y_i,z_i,n_i)u_i \tag{14}$$

where,

$$\begin{aligned}f_i(y_{i-2},y_{i-1},y_i,y_{i+1},y_{i+2},z_{i-1},z_i,z_{i+1})&=c_{1i}^y g_{i-1}(y_{i-2},y_{i-1},y_i,z_{i-1})+(c_{2i}^y-\partial R_y(y_i,z_i)/\partial y_i)g_i(y_{i-1},y_i,y_{i+1},z_i)+\\ &c_{3i}^y g_{i+1}(y_i,y_{i+1},y_{i+2},z_{i+1})-(\partial R_y(y_i,z_i)/\partial z_i)\{c_{1i}^z z_{i-1}+c_{2i}^z z_i+c_{3i}^z z_{i+1}-R_z(y_i,z_i)\}\end{aligned} \tag{15}$$

and,

$$b_i(y_i,z_i,n_i)=m.(\partial R_y(y_i,z_i)/\partial z_i).n_i\,. \tag{16}$$

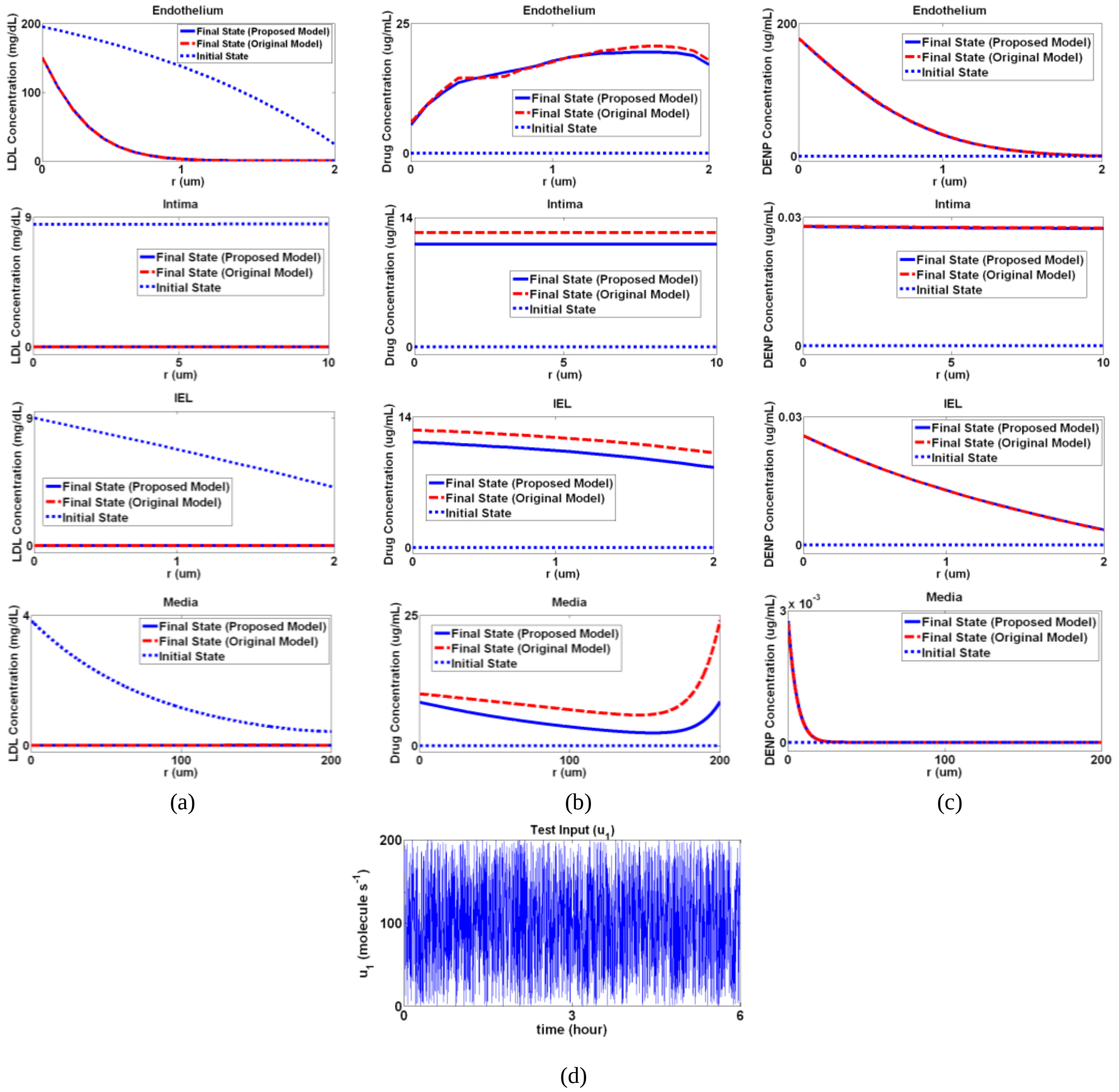

**Figure 2| Comparing the proposed lumped model with the original distributed model through numerical simulation over six hours. a,** LDL concentration profile. **b,** Drug concentration profile. **c,** SPDENP concentration profile. **d,** Test input (only $u_1$ is shown).

## *3.2. Sliding-based Proportional Controller Design*

Here, we apply the concept of sliding-mode control theory to the above proposed lumped model in (14) to design a simple nonlinear SPDENP controller. The proposed controller is basically a biased proportional controller, in which a nonlinear term is added to increase the smoothness of the controller. We use the sliding-mode control theory [25] to prove this controller's stability; hence it is called a sliding-based proportional controller.

From the viewpoint of an SPDENP at the $i^{th}$ finite element node, the goal of control is reducing $y_i$ by manipulating $u_i$ according to the governing system dynamics in (14). From the sliding mode control theory, we define the following sliding surface:

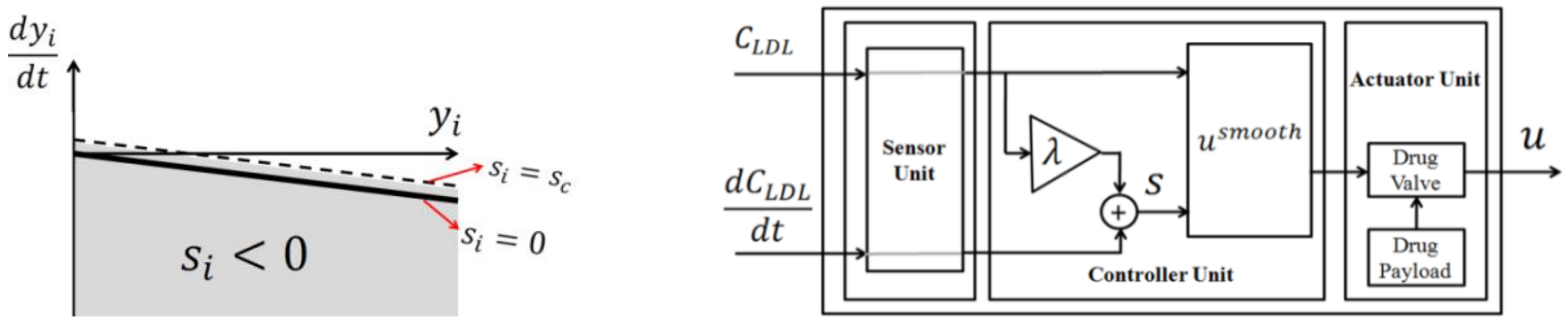

(a) (b)

**Figure 3 | a,** The sliding surface: $s_i = \frac{dy_i}{dt} + \lambda y_i$ . **b,** The structure of SPDENP.

$$s_i = (\frac{d}{dt} + \lambda)(y_i - 0) = \frac{dy_i}{dt} + \lambda y_i \quad (17)$$

where $\lambda > 0$.

Fig.3-a depicts the sliding surface. As displayed in this figure and from (7), $y_i \geq 0$, and thus if the system state is located in the region of $s_i \leq 0$, $y_i$ continuously reduces to zero (because $dy_i/dt < 0$). Hence, the goal of control is to make $s_i$ non-positive. We apply the commonly-used quadratic Lyapunov function $s_i^2$. If the following condition holds, the system state always moves from $s_i > 0$ to $s_i \leq 0$ and remains there:

$$\frac{1}{2}\frac{ds_i^2}{dt} \leq -\eta_i s_i \qquad if \qquad s_i > 0 \quad (18)$$

where $\eta_i > 0$. Now, we determine $u_i$ such that the above inequality always hold. Since $\frac{1}{2}\frac{ds_i^2}{dt} = s_i \frac{ds_i}{dt}$, (18) simplifies to:

$$s_i \frac{ds_i}{dt} \leq -\eta_i \qquad if \qquad s_i > 0 \quad (19)$$

According to (17):

$$\frac{ds_i}{dt} = \frac{d^2 y_i}{dt^2} + \lambda \frac{dy_i}{dt} \quad (20)$$

Substituting (11) and (14) in (20) results:

$$\frac{ds_i}{dt} = f_i + \lambda g_i - b_i u_i \quad (21)$$

Using (21) in (19) leads to:

$$f_i + \lambda g_i - b_i u_i \le -\eta_i \quad if \quad s_i > 0 \tag{22}$$

According to (16), $b_i$ is always non-negative. Hence, rearranging (22) gives the following lower bound for $u_i$ (for $b_i \neq 0$):

$$u_i \ge \frac{1}{b_i}(\eta_i + f_i + \lambda g_i) \quad if \quad s_i > 0 \tag{23}$$

The upper bounds of $f_i$ and $g_i$ can be accurately approximated by considering the effect of uncertainty through simulation and presented as $\bar{\bar{f}}_i$ and $\bar{\bar{g}}_i$, respectively. Similarly, the upper bound of $\lambda$ in the hardware implementation is assumed to be $\bar{\bar{\lambda}}$. It should be noted that in the simulation section of this paper $\bar{\bar{\lambda}} = 1.2\lambda$.

According to (16), if $n_i$ is about zero, $b_i$ is also about zero. Since $b_i$ is in the denominator of (23), this would imply that the corresponding $u_i$ in (23) is extremely large and cannot be generated in practice. Fortunately, few minutes after releasing SPDENPs in the blood flow ($\hat{C}_{DENP}$), the concentration of SPDENPs in the interior of the arterial wall ($n_i$) begins to increase and reaches its steady state value ($n_i^{ss}$). This fact can be mathematically proven by (9). Since $A_n$ is Hurwitz, the state equation in (9) is stable and generates a steady state response $n_i^{ss}$ with respect to $\hat{C}_{DENP}$. Thus, we design $u_i$ according to the steady state value of $b_i$, i.e. $b_i^{ss}$, that is defined as follows:

$$b_i^{ss} = m \left.\frac{\partial R_y}{\partial z}\right|_{(0.1\alpha_i, 0.1\beta_i)} n_i^{ss} \tag{24}$$

As a result, (23) is converted to the following inequality:

$$u_i \ge \frac{1}{b_i^{ss}}(\eta_i + \bar{\bar{f}}_i + \bar{\bar{\lambda}}\bar{\bar{g}}_i) \quad if \quad s_i > 0 \tag{25}$$

If $u_i$ satisfies the above inequality, $s_i$ is guaranteed to reduce to zero at the steady state in the presence of uncertainty.

Since SPDENP is designed for patients with very high LDL levels, $C_{LDL}$ in lumen is greater than 200 mg/dL [21]. To prevent drug release in the lumen, $u_i$ is set to zero for $C_{LDL} \ge 160$ mg/dL. Following this assumption, let us consider $u_i$ as follows:

$$u_i = (1-\frac{y_i}{160})u_c \quad if \quad s_i > 0 \quad and \quad y_i < 160\tfrac{mg}{dL} \tag{26}$$

where $u_c$ is a positive constant and satisfies (25) as follows:

$$(1-\frac{\bar{\bar{y}}_i}{160})u_c \geq \frac{1}{b_i^{ss}}(\eta_i + \bar{\bar{f}}_i + \bar{\bar{\lambda}}\bar{\bar{g}}_i) \quad if \quad s_i > 0 \quad and \quad y_i < 160 \tag{27}$$

where $\bar{\bar{y}}_i$ is the upper bound of $y_i$ that can be accurately approximated through simulation in the presence of uncertainty. If $y_i$ is about 160 mg/dL, the condition $y_i < 160$mg/dL in (26) makes $u_i$ non-smooth especially in the presence of noise, in which $u_i$ constantly switches between 0 and a large nonzero value. The term $(1-\frac{y_i}{160})$ in (26) is used to improve the smoothness of control output. Regarding (26), $u_c$ is proportional to $u_i$ and should be small enough to be practically realizable on SPDENP. $b_i^{ss}$ is very small in Media and this leads to a considerable increase in the value of $u_c$ in (27). To address this problem, since LDL level is very low in Media, only the finite element nodes of the first quarter of Media are included in calculating $u_c$ in (27). As $dy_i/dt$ and $y_i$ are measured by nanosensors, these values and consequently $s_i$ are noisy in practice. Thus, the condition $s_i > 0$ in (26) causes chattering around $s_i = 0$.

To improve the smoothness of controller output, we are inspired by sliding mode control theory and modify (26) as follows:

$$u_i^{smooth} = \begin{cases} (1-\frac{y_i}{160})u_c & s_i > s_c \quad and \quad y_i < 160\frac{mg}{dL} \\ (1-\frac{y_i}{160})(\frac{s_i}{s_c})u_c & 0 \leq s_i \leq s_c \quad and \quad y_i < 160\frac{mg}{dL} \\ 0 & otherwise \end{cases} \tag{28}$$

where $s_c$ is a positive constant and determines the width of marginal layer, $0 \leq s_i \leq s_c$, as shown in Fig.3-a. Following (26) and (27), the first line of (28) states that $ds_i/dt$ is necessarily negative in the presence of uncertainty when $s_i > s_c$. This implies that the resulting controller (28) always pushes the system state into $s_i \leq s_c$.

From the above analysis, the final proposed controller (28) is basically a biased proportional controller $(1-\frac{y_i}{160})u_c$, in which a nonlinear term $(1-\frac{y_i}{160})(\frac{s_i}{s_c})u_c$ is added to increase smoothness. It should be noted that large values of $u_c$ enhance the robustness of the controller in the presence of uncertainty. Also, large values of $u_c$ , $\eta_i$, and $\lambda$ increase the controller speed. But if the values of these design parameters are large, it considerably increases drug consumption, which is undesirable in practice. Thus, there is a trade-off between robustness, speed, and drug consumption.

### *3.3. Structure of SPDENP*

Fig.3-b shows the structure of SPDENP consisting of sensor, controller, and actuator units. In this figure, $C_{LDL}$ and $dC_{LDL}/dt$ are the LDL concentration and its rate of change that are sensed by the nanoscale molecular concentration sensors of SPDENP from the environment. $\lambda$ and $s$ are defined in (17), $u_{smooth}$ is described in (28), and $u$ is the drug release rate of SPDENP. The actuation unit of SPDENP contains a controllable drug valve connected to the drug payload that releases the drug molecules in the aqueous environment with a flow rate determined by the control unit.

In this study, it is assumed that SPDENP has a limited lifetime (above few days). Biodegradability is considered as the assumed clearance mechanism of SPDENP. After the lifetime of SPDENP is finished (even if SPDENP is still full of the drug), SPDENP is broken into small biocompatible molecules (similar to the waste products of metabolism) and cleared from living tissues through the natural clearance mechanism of the human body. In this paper, we consider a nanoparticle as a system, and our insight to SPDENP is abstract and mathematical. Thus, SPDENP introduces an abstract model for a nanoparticle, not a specific nanoparticle with certain material and physical-chemical structure. Accordingly, we assume that the SPDENP is composed of biocompatible materials, especially after biodegradation. We should add that the proposed SPDENP exploits the natural diffusion/advection for its locomotion, hence it does not require a propulsion unit.

### *3.4. Experimental Feasibility*

Remarkable experimental works have been performed in recent years on constructing SPDENP's various nanoscale components, such as its molecular concentration sensors, a nanoscale computing unit, and a nanometer-sized controllable valve. Below, we briefly review several of these related experimental works.

Nanosensors have been widely used for biological and medical applications in the literature [26]. Many nanosensors have been designed based on advanced nanostructures, such as smart multifunctional nanoparticles for drug delivery [27], dual-ligand functionalized carbon nanodots for cellular targeted imaging [28], functionalized nanotubes for in-vivo pharmacological measurements of a chemotherapy drug [29], and mesoporous silica nanoparticles for sensitive detection of a psychedelic drug [30]. Also, biotechnology have been employed for this goal such as DNA-based [31] and aptamer-based [32] nanosensors. There exist different technologies for designing nanoactuators including stimuli-responsive nanovalves [33], mechanosensitive channel [34], mesoporous silica nanoparticles [35], optogenetics [36], and DNA-based nanoactuators [37]. In case of realizing nanoscale in-vivo computing units, DNA technology is the most common approach such as the logic gates based on DNA origami for universal computing [38], programmable chemical controllers made from DNA [39], and molecular simple logic programs [40].

Although these smart therapeutical nanoagents have been used in different diseases, most researches have been focused on cancer therapy [41-43]. Authors in their prior researches [17-19] also proposed the theoretical notion of 'swarm control systems for nanomedicine' in order to realize swarm intelligence by simple therapeutical nanoagents. By this scheme, the concept of intelligent drug delivery is developed for treating atherosclerosis [17-19] and cancer [44]. Since nanoagents need to communicate with each other, various signaling methods have been introduced for realizing nano-communication networks [45]. In the context of nanoscale drug delivery, a lot of researches have been focused on molecular and electromagnetic communication [46-47].

While the size of components in some of the above works is larger than 100 nm (the desirable size of SPDENP), the above-mentioned advances in recent years carry a considerable promise that even smaller nanoscale components can be realizable in the near future. Admittedly, there are remaining practical challenges to the realizability of SPDENPs. Yet, as the present study suggests, SPDENPs carry a simpler mathematical structure when compared with many other competing strategies. This leads to a more reasonable experimental realizability and manufacturing of smart nanoparticles such as those proposed in this study.

### *4. Simulation Results*

In this section, a swarm of the proposed SPDENPs is employed to control the LDL level in the arterial wall. The original mathematical model (equations (1)-(3)) is simulated in MATLAB. Mesh size, time step, lumen concentrations of LDL, SPDENP, drug, and their initial levels in the arterial wall are similar to Section 2-c. In this simulation, $R_y(y_i, z_i) = k_{rLDLdrug} y_i z_i$ , $R_z(y_i, z_i) = k_{rdrugLDL} y_i z_i$ ,

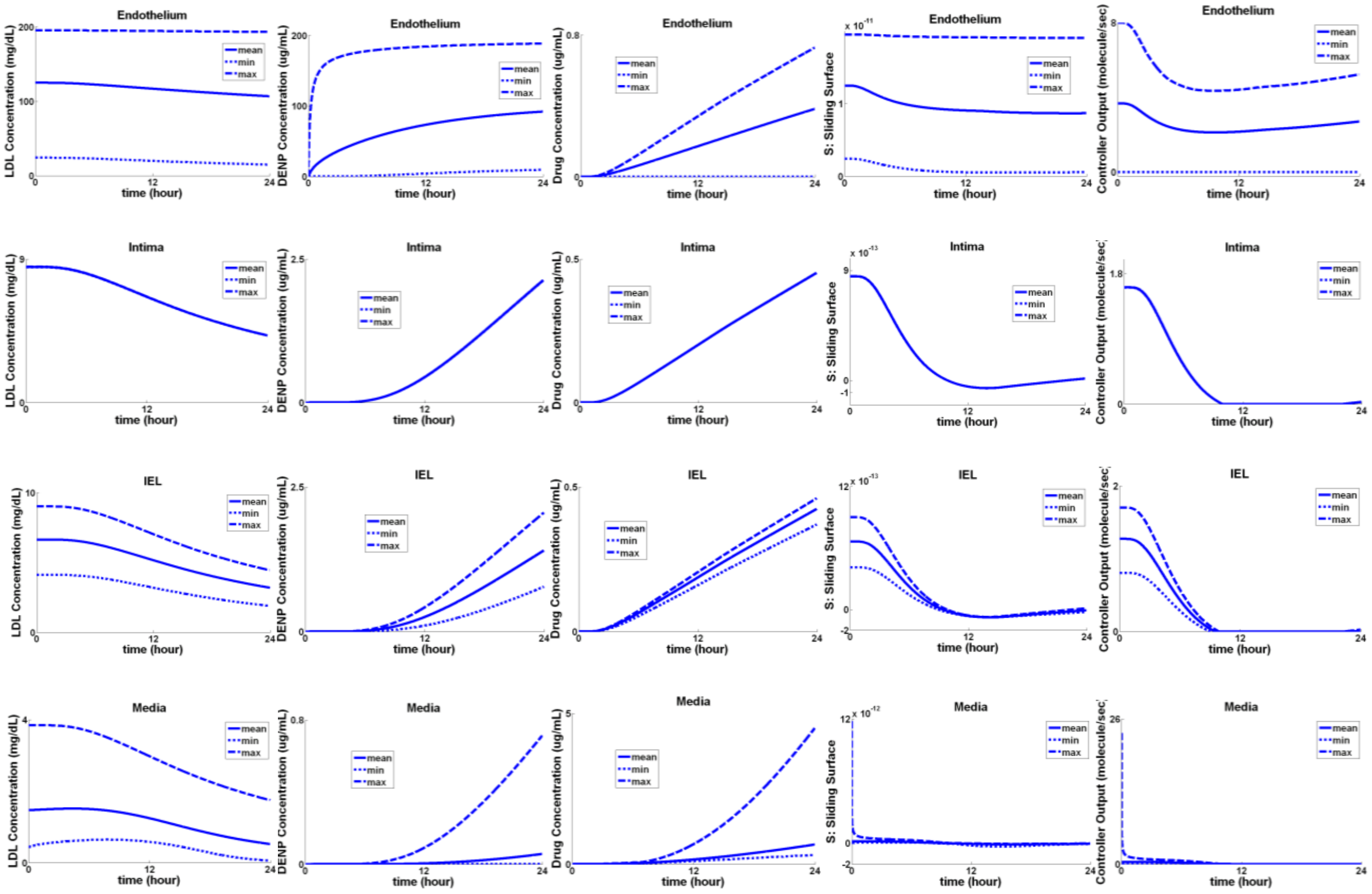

(a) (b) (c) (d) (e)

**Figure 4 | The effect of the proposed SPDENP on an *unhealthy* (*abnormal)* arterial wall (*LDL level of* 200 mgdL$^{-1}$) over one day. a,** LDL concentration. **b,** SPDENP concentration. **c,** Drug concentration. **d,** Sliding surface. **e,** Controller output, versus time.

$\lambda = 10^{-5}, \eta_i = 10^{-15}$ and $s_c = 10^{-10}$. The reactions coefficients and $m$ are also set according to [18]. The mass of each SPDENP is assumed to be 60000 times larger than $m$. Choosing $u_c$ = 200 molecule/second, Fig.5-c shows that (27) holds for the most critical points of the arterial wall, i.e. from the middle of Endothelium ($y_i < 160$ mg/dL) to the end of the first quarter of Media. This value of $u_c$ (maximum drug release rate) is considerably smaller than its corresponding value in the previous works [18]. Consequently, the required drug release rates can be realized by much simpler drug pumps, which can significantly simplify the implementation of SPDENP.

We consider the effect of SPDENPs on the LDL level of the interior of an unhealthy (abnormal) arterial wall over 24 hours (one day). Fig.4 illustrates the concentration levels of LDL, SPDENP, drug, sliding surface, and controller output over time in each wall layer. In these plots, "min", "mean", and "max" stand for the minimum, average, and maximum value of the related signal over all points of the mentioned wall layer. Fig.4 shows that the diffusion patterns of SPDENP and LDL concentrations are different from those of the drug. SPDENP and LDL concentrations get their largest value in Endothelium and smallest value in Media, while the drug concentration has an opposite pattern. This is due to two reasons. First, the drug concentration's boundary condition in lumen is zero, while the boundary conditions of SPDENP and LDL are non-zero. Second, the direction of fluid velocity ($V_{filt}$) is from lumen to Media. Hence, the concentration of drug gets its largest value in Media.

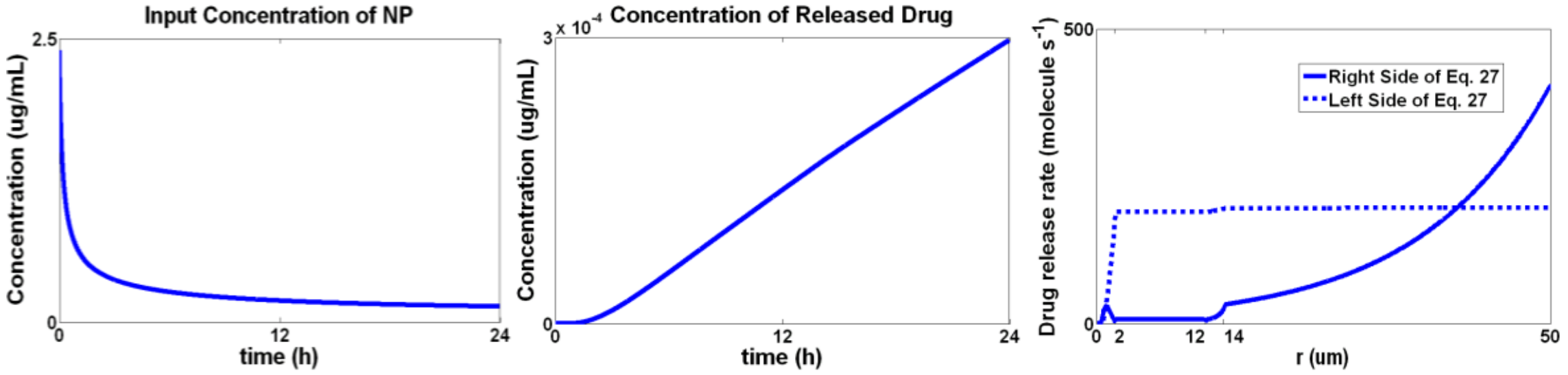

(a) (b) (c)

**Figure 5 | a,** Input concentration of SPDENP from lumen to arterial wall at each time. **b,** Average concentration of the drug released from SPDENPs into the arterial wall at each time. **c,** $u_c$ = 200 molecule $s^{-1}$ satisfies Eq.27 for most of the points of the arterial wall.

In Fig.5, the input concentration of SPDENP from lumen to the arterial wall, and the average concentration of the drug released from SPDENPs in the interior of the arterial wall are shown at each time. Fig.6 depicts the final (controlled) profile of sliding surface and LDL concentration in the arterial wall at the end of 24 hours. Also, Fig.6-b compares the final (controlled) LDL level with the desired (normal) and uncontrolled (without drug) LDL levels. The uncontrolled LDL level is already defined in Section 2-C. The desired (normal) LDL level is defined similarly while the lumen concentration of LDL is 100 mg/dL (corresponding to normal (desirable) LDL level [21]).

Fig.6-b demonstrates that although the LDL concentration in lumen is 200 mg/dL (according to [21], this level of LDL is known as very high LDL level), the proposed SPDENP could successfully reduce the LDL level in all layers of the arterial wall with reduction rate of 14.6%, 50.5%, 51.8%, and 64.4% in Endothelium, Intima, IEL and Media, respectively. Here, the reduction rate is defined as:

$$\%\ \text{improvement} = mean\left(\frac{\text{Uncontrolled} - \text{Controlled}}{\text{Uncontrolled}}\right) \times 100 \qquad (29)$$

Comparing these numbers with the results of the previous work of authors [18], 17.9%, 45.7%, 46.9% and 61.3%, shows that SPDENP has better performance, in contrast to PDENP, in Intima, IEL, and Media layers. Also, the controlled LDL level is even better than the desired value in these three layers. As expected, the controlled LDL level in Endothelium is higher than the desired value. This is due to the controller parameters (the condition $y_i < 160\ \mathrm{mg/dL}$ in (28)) such that it does not release any drug in lumen. Since the range of LDL concentration at the lower points of Endothelium is close to the LDL level of lumen, SPDENP does not release any drug in these points and thus LDL level is not significantly reduced. The area under the curve of Fig.5-b is $0.00334\mu g.h/mL$. Comparing this value with the results of PDENP in [18], $0.00390\mu g.h/mL$, demonstrates that the drug consumption of SPDENP is also more efficient than PDENP.

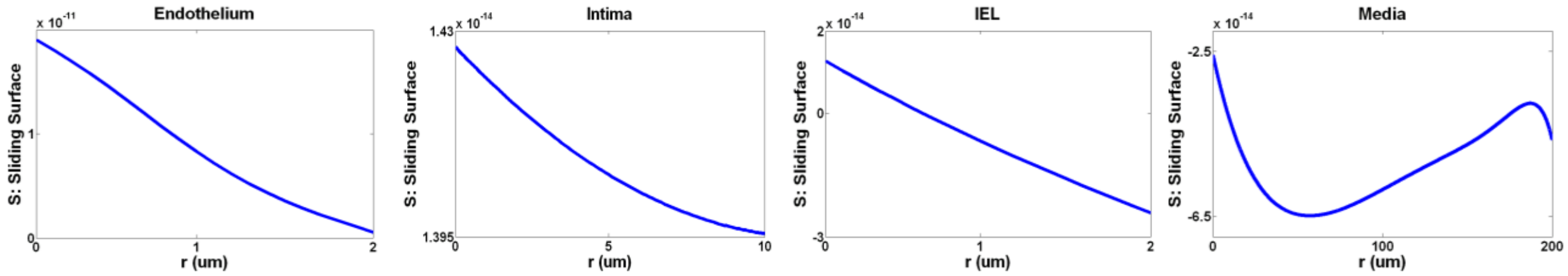

(a)

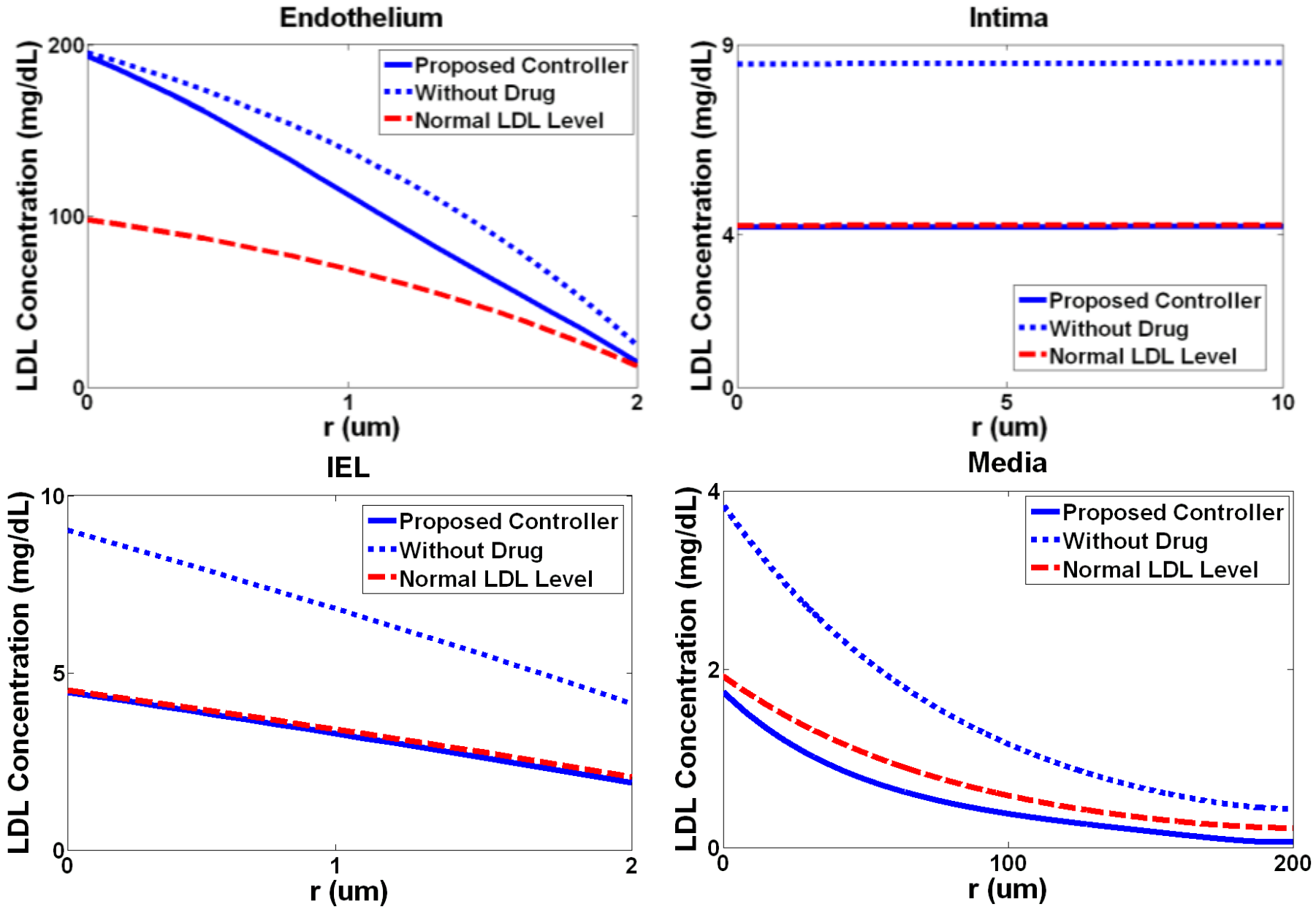

(b)

**Figure 6 | a,** Final profiles of sliding surface after one day. **b,** Final LDL concentration profiles after one day in contrast to normal (desired) LDL level and uncontrolled (without drug) LDL level.

## 5. Conclusion

The point of departure in the present study is introducing the principles of mathematical proof in the design of nanoparticles for targeted drug delivery. In particular, the control systems theory is used to devise a robust and stable swarm approach for the atherosclerosis prevention and treatment. Since the high accumulation of LDL molecules plays a significant role in the initiation and development of atherosclerotic plaques, the control of LDL level within the arterial wall can be a crucial approach to prevent and treat atherosclerosis.

More specifically, we introduce a nonlinear lumped mathematical model for mass transport in the arterial wall and evaluate its accuracy compared to the original distributed model. This finite element lumped model and the concepts of sliding-mode control theory are then applied to analytically design the proposed sliding-based proportional drug encapsulated nanoparticles (SPDENP). In contrast to most of the treatment methods of nanomedicine that are designed empirically, the proposed controller is designed based on mathematical proof, and the overall stability of the system is analytically proven under uncertain dynamics.

Simulation results on a well-known model of the arterial wall demonstrate that although the LDL concentration in lumen is very high, the proposed SPDENPs successfully reduce the LDL level in all layers of an unhealthy arterial wall with a reduced rate of 14.6%, 50.5%, 51.8%, and 64.4% in Endothelium, Intima, IEL and Media, respectively. Also, comparative analysis shows that both LDL reduction rate and drug consumption of SPDENP are generally improved with respect to PDENP. It should be mentioned that while the proposed SPDENP is specifically designed here for the prevention and treatment of atherosclerosis, the general approach is similarly applicable to other diseases where targeted drug delivery can be deemed effective. The notion of SPDENP is generally an applicable method for realizing robust control at nanoscale, particularly when the presence of uncertainty significantly affects the complexity of system behavior.